\documentclass[12pt]{iopart}
\usepackage{graphicx}
\usepackage{amsthm,amssymb,amscd}

\begin{document}

\title[Effect of momentum deposition on flow anisotropy]{%
The effect of momentum deposition during fireball evolution on flow anisotropy}

\author{Martin Schulc$^{1,2}$ and Boris Tom\'a\v{s}ik$^{1,3}$}

\address{$^1$Czech Technical University in Prague, FNSPE, 11519 Prague, Czech Republic}
\address{$^2$Research Centre \v{R}e\v{z}, 25068 Husinec-\v{R}e\v{z}, Czech Republic}
\address{$^3$Univerzita Mateja Bela, Fakulta pr\'irodn\'ych vied, 
97401 Bansk\'a Bystrica, Slovakia}
\eads{\mailto{martin.schulc@cvrez.cz}}

\begin{abstract}
The highest LHC energies give rise to  production of many pairs 
of hard partons which deposit 
four-momentum into the expanding fireball matter. 
We argue that it is necessary to include momentum deposition during fireball evolution 
into 3+1 dimensional hydrodynamic simulations of the collision.
This influence cannot be accounted for simply by modifying initial conditions of  the 
simulation. The resulting contribution to flow anisotropy is correlated with the 
fireball geometry and causes an increase of the elliptic flow in non-central collisions. The results are presented for various scenarios with energy and
momentum deposition to clearly demonstrate this effect.
\end{abstract}



\section{Introduction}
\label{s:intro}
The goal of  high-energy heavy-ion collisions is to study properties of  
matter in deconfined state. Relativistic 
hydrodynamics has proven to be very successful tool for modelling the bulk 
dynamics of the hot dense matter formed during a heavy 
ion collision \cite{Song, Nonaka, Huovinen:2001, Schenke:2010, Huovinen:2005, Hirano}. Comparisons of hydrodynamic simulations with 
the measured data aim at extracting the equation of state (EoS) and 
the transport coefficients, such as the  shear and bulk viscosities. 
The importance of including 
fluctuations into hydrodynamic calculations has been recognized, 
leading to a new set of experimental observations of higher flow coefficients 
and their correlations \cite{Qiu, Schenke:2011, Alver, Teaney}. Event-by-event viscous hydrodynamic modelling has proven to be essential for correctly 
describing all the details of the bulk behaviour of heavy ion collisions. 
The standard approach being used now is to select  
a set of initial conditions and to tune the values of transport coefficients
in order to find such a setting of hydrodynamic simulations 
which reproduces all features of the data. Simulations indicate  
linear mapping of initial state spatial anisotropies  onto final state momentum 
distribution anisotropies on event-by-event basis \cite{gardim:12,niemi:13,gale:13}. 
The mechanism, that is being investigated here, breaks such a direct correspondence. 

It has been demonstrated that at the LHC energy, momentum deposition 
from hard partons into expanding matter effectively produces measurable anisotropy 
in transverse expansion \cite{jets,Tachibana:2014lja,Andrade:2014swa}. 
Here we show that the effect cannot be simulated by just 
including momentum anisotropies into initial conditions. 
Jets and minijets are produced copiously in initial hard scatterings 
at the highest achievable energies at the 
LHC and propagate through the deconfined medium. The quark-gluon plasma extensively quenches the energy and momentum carried by hard partons which might become jets. 
Their energy loss when traversing the matter is so huge that only a few of them appear as distinguished jets \cite{Adams, Aad, Chatrchyan, Aamodt}. The momentum and 
energy depositions from partons into bulk medium induce collective phenomena there \cite{Satarov, Solana, Koch}.
The approach presented in this paper bears two distinctive features. 
Firstly, energy loss is accompanied by deposition of momentum directed along the hard parton 
motion. 
Secondly, it is being deposited over some period of time and not just instantaneously at the beginning of the expansion. 

In our previous work \cite{jets} we have shown that the interplay of many 
hard-partons-induced 
streams in a single nuclear collision at the 
LHC yields considerable contribution to azimuthal anisotropies of hadron distributions. 
In non-central collisions the contribution is correlated with fireball geometry. 
Surprisingly,  it was found that the magnitude of this effect does 
not significantly depend on the value of ${dE}/{dx}$, at least for the two tested values of the 
energy loss. 
This triggers the question whether the effect on flow anisotropies would not be the same 
if all energy and momentum were deposited in the initial conditions. The argument favouring 
such a conclusion might be that the total amount of deposited energy and momentum is always
the same, and a  higher value of $dE/dx$ only means that it is deposited faster. 
The case which includes energy and momentum deposition in initial conditions actually 
corresponds to an infinite value of ${dE}/{dx}$. 
In the present paper
it will be compared  with scenarios where energy-momentum is deposited during 
evolution of the fireball. 
We show that including momentum and energy deposition only in the initial stage 
is not enough to create azimuthal anisotropy of the same size as in case with deposition 
during the evolution of the fireball. 

We also investigate more fine-tuning details of the scenario with energy and momentum deposition 
\emph{during} the hydrodynamic evolution. Our conclusions appear robust against reasonable changes  
of the model parameters.

The paper is organised as follows. Section \ref{s:method} 
introduces the methods and ideal three-dimensional hydrodynamic 
model that was used in our simulations. 
Then, Section \ref{s:res} presents our results obtained with this model for the 
simulation of energy and momentum deposition from pairs of hard partons 
in different configurations. 
Important conclusions are summarised in Section \ref{s:summ}.


\section{Calculation methods}
\label{s:method}

To simulate the energy-momentum deposition from hard partons at the beginning 
or during the evolution of the fireball we have employed ideal 
event-by-event three-dimensional hydrodynamic model. It uses the SHASTA 
algorithm \cite{Boris, DeVore} to treat strong gradients and shocks without dispersion.

The initial conditions are calculated by means of smooth optical Glauber prescription. We have chosen these simple initial conditions since any contribution of 
hard partons generates distinct features over this background and 
can be easily distinguished. By using different initial conditions as \textit{e.g.}\
Glauber Monte Carlo, distinction of hard partons 
effect would be very hard  or almost impossible.
The nucleon-nucleon cross-section for Glauber calculation at $\sqrt{s_{NN}} = 5.5$~TeV 
was set to 62 mb. 
The shape of  initial energy density distribution in the transverse plane is parametrized as 
\begin{equation}
\label{e:w}
W(x, y; b) = (1-\alpha)n_{w}(x, y; b) + \alpha n_{bin}(x, y; b), 
\end{equation}
where $n_{w}$ and $n_{bin}$ are the numbers of wounded nucleons and binary collisions at given transverse position $(x, y)$ and the coefficient $\alpha$ which 
determines the fraction of the  binary collisions contribution was set to 0.16. 
For the longitudinal profile we employed the prescription used in \cite{Nonaka, Music}. 
It consists of two parts, a flat region around $\eta_{s}$ = 0 and half-Gaussian 
in the forward and backward end of the plateau:
\begin{equation}
\label{r:eta}
 H(\eta_{s}) = \exp \left[-\frac{(|\eta_{s}| - \eta_{flat}/2)^{2}}{2\sigma_{\eta}^{2}}\theta(|\eta_{s}| - \eta_{flat}/2) \right].
\end{equation}
The parameters $\eta_{flat}$ = 10 and $\sigma_{\eta}$ = 0.5 were chosen according to \cite{Schenke}. The complete energy density distribution is then given by
\begin{equation}
\label{e:all}
\epsilon(x, y, \eta_{s}, b) = \epsilon_{0} H(\eta_{s}) \frac{W(x, y, b)}{W(0, 0, 0)}\, .
\end{equation}
We choose $\epsilon_{0}$ = 60 GeV/fm$^{3}$ and the 
initial proper time $\tau_{0}$ was set to 0.55 fm.

The initial expansion velocity field was longitudinally boost invariant without any transverse component. 
This was modified only in the hotspots-with-momentum model, as we explain below. 

To close the hydrodynamic set of equations we employed a lattice-inspired nuclear equation of state \cite{Petreczky}. 

Incorporation of energy and momentum deposition due to 
hard partons into hydrodynamic equations was done via source terms 
in the energy-momentum conservation equation
\begin{equation}
\partial_{\mu} T^{\mu \nu}(x)=J^{\nu}, \label{eq:energy} 
\end{equation}
where $T^{\mu \nu}(x)$ is the energy-momentum tensor and the 
term $J^{\nu}$ on the right-hand side is the parametrised source 
term describing deposition of energy and momentum into the medium \cite{Betz1, Betz2} 
\begin{equation}
\label{e:jnu}
J^{\nu}=-\sum_{i} \int^{\tau_{f},i}_{\tau_{i}, i}d\tau \frac{d P^{\nu}_{i}}{d \tau}\delta^{(4)}(x^{\mu}-x^{\mu}_{jet,i}),
\end{equation}
where $x^{\mu}_{jet,i}$ denotes the position 
of $i$-th hard parton and  $dP^{\mu}_{i}/d\tau$
is its momentum change. Integration runs over the whole lifetime 
of $i$-th parton until its energy is fully deposited and the summation goes 
over all hard partons of the event. 
Their source term is in non-covariant notation implemented 
in a form of a three-dimensional Gaussian function in $x$, $y$ and $\eta$ coordinates \cite{jets, twojets,Betz1}
\begin{equation}
J^{j} = -\sum_i \frac{1}{(2\, \pi\, \sigma_i^2)^{\frac{3}{2}}} \, \exp \left (
- \frac{\left ( \vec x - \vec x_{{jet},i} \right )^2 }{2\, \sigma_i^2} \right )\, 
\left ( \frac{dE_i}{dt}, \frac{d\vec P_i}{dt} \right )\, .
\label{e:jgauss}
\end{equation}
The sum runs over all the jets and the Gaussian profiles are centered around the actual positions of the partons.
The width of the Gaussians $\sigma_i$ was set to 0.3 fm as in \cite{jets}. 
We have checked that tuning the width to 0.15~fm or to 0.6~fm does not lead to major change of our results.  

Since it is more suitable for the nature of the problem, hydrodynamic simulations have been performed in Milne coordinates
$\eta = \frac{1}{2}\ln((t+z)/(t-z))$ and $\tau= \sqrt{t^2 - z^2}$. The formulation of the source term has been adjusted accordingly in 
the numerical procedure. They are evolved in proper time, as indicated in Eq.~(\ref{e:jnu}). The widths which appear
in Eq.~(\ref{e:jgauss}) refer to Cartesian coordinates and cannot be used for $\eta$ without any change. 
Indeed, in order to keep  constant longitudinal width for the energy deposition, the width in $\eta$ has been 
scaled by $\tau^{-1}$.

Parton energy loss depends on the density of the surrounding medium. We assume that parton energy loss scales with entropy density as in \cite{Betz3}.  The scaling relation is then
\begin{equation}
\frac{dE}{dx} = \left . \frac{dE}{dx}\right |_0 \,  \frac{s}{s_0} 
\end{equation}
with $s_0$ corresponding to energy density 20.0~GeV/fm$^3$ 
($T=324$~MeV with the used EoS).  This gives $s_0=78.2/\mathrm{fm}^{3}$.
For $dE/dx|_0$ we usually choose the value 7~GeV/fm in this paper. 
We have also made simlations, however, with very low $dE/dx$ in order to study the 
limit of vanishing energy loss. 

For the production of hard partons we take the parametrisation of gluon cross-section per nucleon-nucleon pair in nucleus-nucleus collisions from \cite{toymodel}
\begin{equation}
 E \frac{d \sigma_{NN}}{d^{3}p}=\frac{1}{2\pi} \frac{1}{p_{t}}\frac{d \sigma_{NN}}{d p_{t} dy}=\frac{B}{(1+p_{t}/p_{0})^{n}},
\end{equation}
where for the energy $\sqrt{s_{NN}}=5.5\,\mathrm{TeV}$ we have $B= 14.7$~mb/GeV$^{2}$, $p_{0}=6$ GeV and $n=9.5$. 
This equation describes the initiation of hard partons in our hydrodynamic model at $\tau_{0}$.
Pairs of hard partons are generated back-to-back in transverse momentum with directions chosen in  
azimuthally symmetric manner. 
Thus the transverse momentum is always conserved for each pair of hard partons. 
Rapidities of the hard partons are generated from a uniform distribution and they are
chosen independently for both partons of a pair. 

The starting distribution of hard parton pairs in transverse plane scales 
with the number of binary collisions. 
Initial positions in $\eta$ are chosen from uniform distribution. 
For the presented results we generated partons with $p_{t}$ above 3 GeV.
Altogether this procedure spits in average about 10 hard partons per central event into our simulation.
More details of the whole procedure can be found in \cite{jets,toymodel}.

Recorded freeze-out hypersurfaces given by $T = $ 150 MeV were processed by the Cooper-Frye formula \cite{Cooper}. 
For every scenario recorded below we generated 100 hydrodynamic 
events. We use the THERMINATOR2 package \cite{THERMINATOR2} 
to generate observable hadrons on the obtained hypersurface and evaluate the results. 
For every hydrodynamic event there were five THERMINATOR2 events 
generated, thus giving 500 
events in total for each setting. Resonance decays were 
included. Anisotropic flow parameters $v_{1}, v_{2}, v_{3}, v_{4} $ for charged hadrons were extracted by the two-particle 
cumulant method.


\section{Results}
\label{s:res}

The philosophy of our approach is to present the effect of any source of momentum 
anisotropy against the baseline given by simulation with smooth non-fluctuating 
optical Glauber initial conditions. Thus we always show such results which are marked 
``no-jets''. We do so even if they are trivial as e.g.\ in the case of central collisions, 
also in order to validate our hydrodynamic model. 

It is currently an open question, what is the value of $dE/dx|_0$. Surprisingly, in 
\cite{jets} we observed that the anisotropic flow coefficients did not change whether 
we chose 4~GeV/fm or 7~GeV/fm.  Hence, it is interesting to investigate 
the limit case where $dE/dx|_0 \to \infty$. 
It corresponds to the case where momentum and energy is deposited in initial conditions.
Exploration of this limiting case is important.
If it leads to the same results as the simulations with finite $dE/dx$, then all inhomogeneities 
in energy and momentum density can be put into initial conditions. As a consequence, 
we would recover the linear relationship between initial state spatial anisotropy 
and final state distributions of hadrons \cite{gardim:12,niemi:13,gale:13}.
If, on the other hand, energy and momentum deposition \emph{during} 
the evolution leads to different results, then it must be properly included in 
phenomenologically relevant simulations. 

Moreover, we considered the opposite limit, \textit{i.e.}\ 
$dE/dx|_0 \to 0$. In our simulations, the value of $dE/dx$ was set to 0.1 GeV/fm. 
The results for all anisotropic flow coefficients were found to be compatible with zero. 
The main reason is that the hydrodynamic simulation ends before significant amount of energy or momentum is deposited.

A little more on the technical side, 
we also investigated influence of varying the width of the deposited energy and momentum Gaussian for hard partons scenario. 
The change of the Gaussian width to 0.6 fm gives slightly higher values of anisotropic flow coefficients than results for 0.3 fm. Conversely, 
Gaussian width of 0.15 fm gives slightly lower values of flow coefficients. 
However, the results for all considered widths can be regarded approximately same within the error bars.

\begin{figure}[b]
\centerline{
  \includegraphics[width=0.5\textwidth]{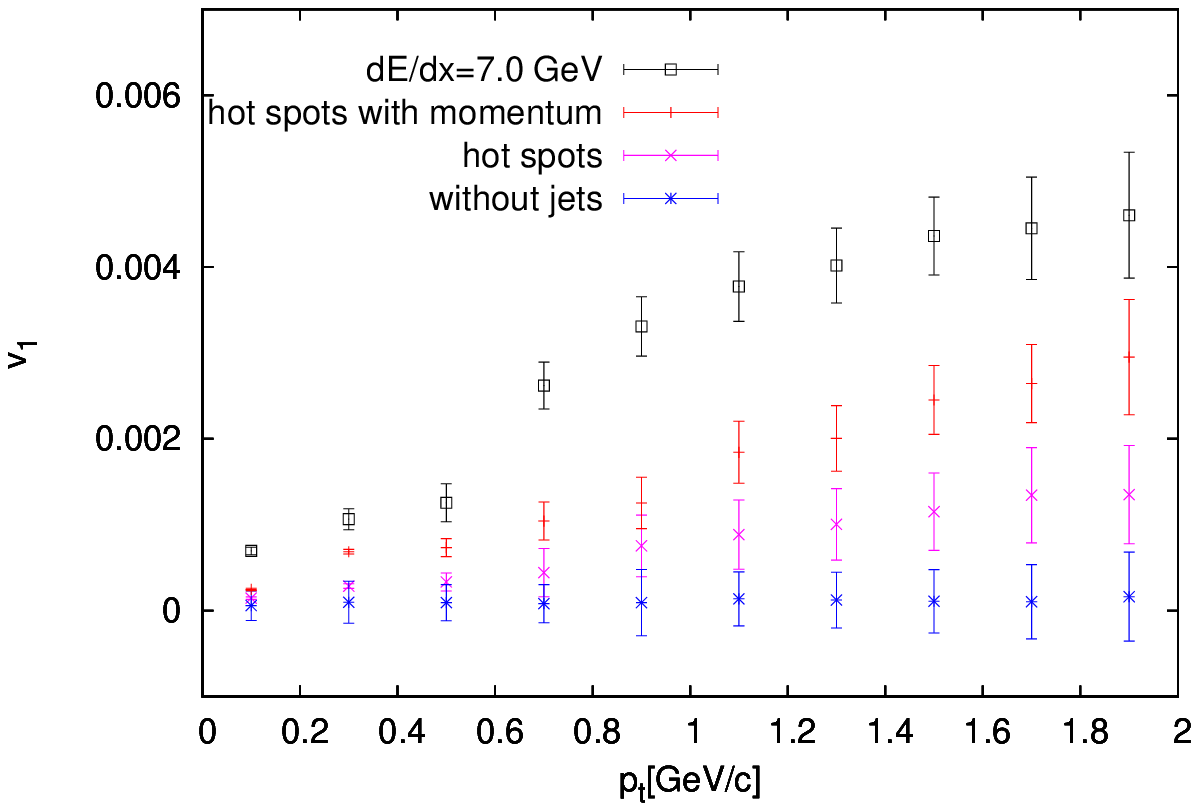}
  \includegraphics[width=0.5\textwidth]{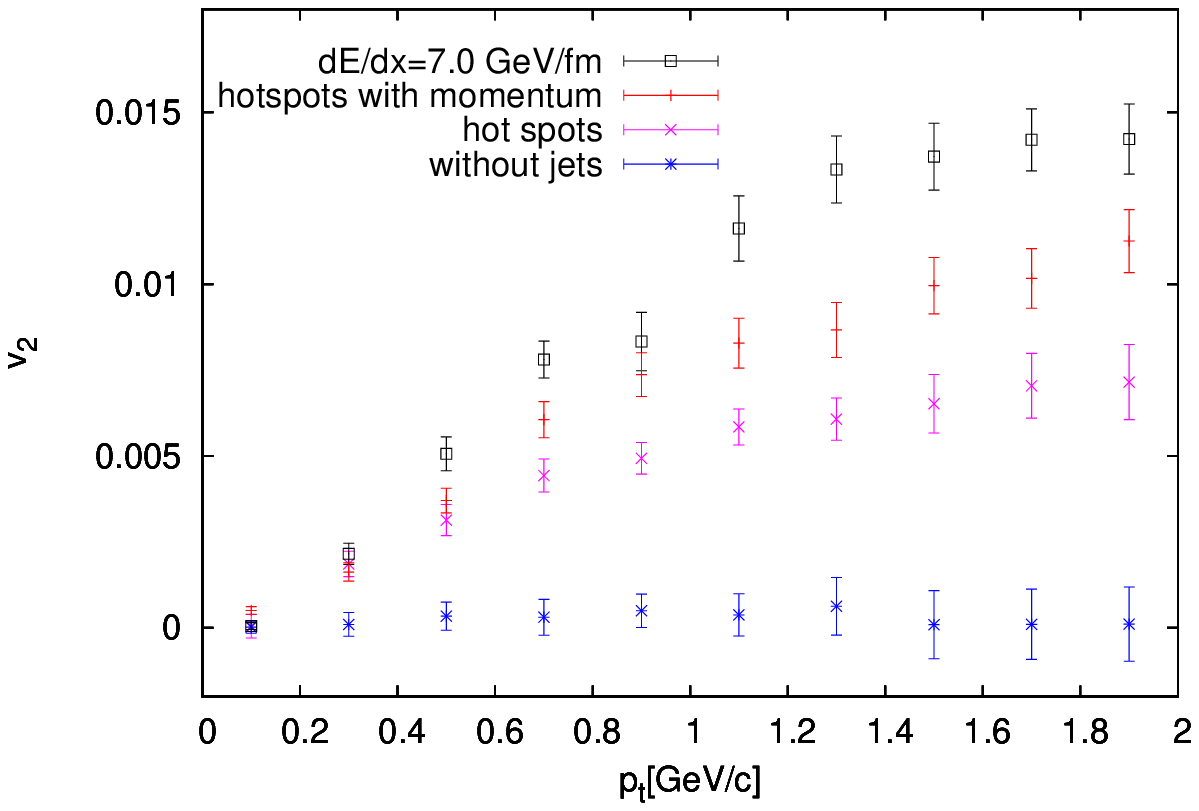}
}
\centerline{
  \includegraphics[width=0.5\textwidth]{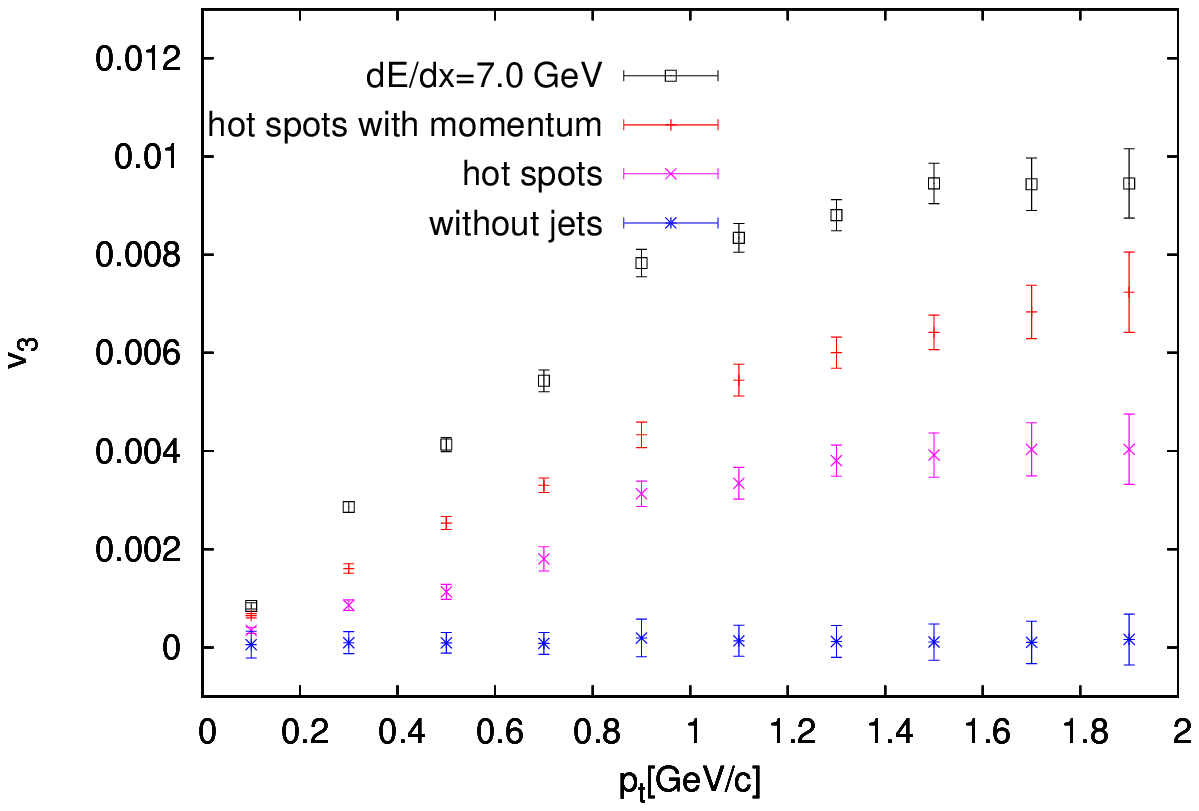}
  \includegraphics[width=0.5\textwidth]{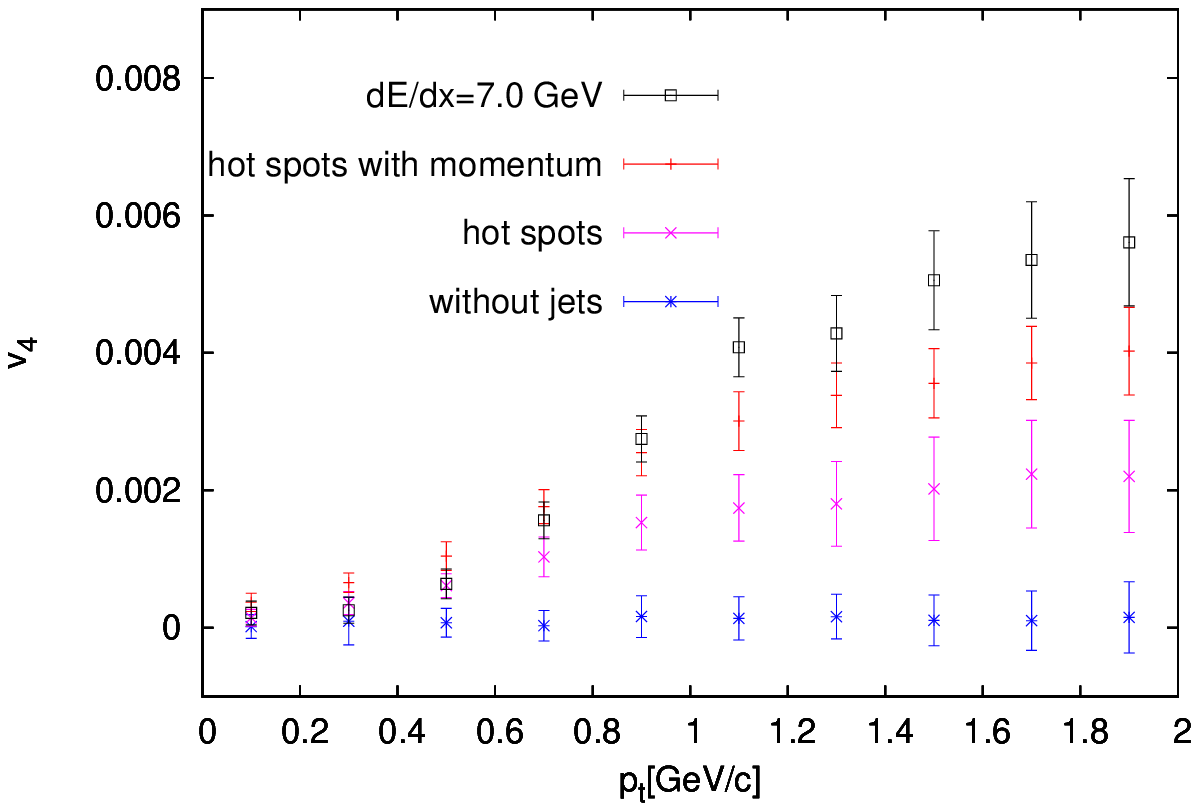}
}
\centering
\caption{Parameters $v_{n}$ from collisions at $b=0$ for charged hadrons. Different symbols represent: 
energy loss of hard parton $dE/dx|_0 = 7$~GeV/fm (black $\square$), 
scenario with hot spots with momentum deposition in initial conditions (red --),
scenario with only hot spots in initial conditions (purple $\times$), 
and scenario with smooth initial conditions (blue *). }
  \label{fig:fig1}
\end{figure}

For our main investigation, we
begin with the study of  the generated anisotropy of momentum distribution 
in ultra-central collisions 
($b=0$) for various scenarios. 
Results are shown in Figure \ref{fig:fig1}. 
Results of our model are represented by a set of ``jet events''  
with $\left. \frac{dE}{dx} \right|_0 = 7$~GeV/fm. 
As a benchmark test we also evaluate the $v_{n}'$s 
from simulation with no hard partons and no fluctuations of the initial state 
and show that they are consistent with 0. 

In order to see if comparable results can be obtained with appropriately set initial 
conditions, we compare the results with 
simulations with hot-spots. Two versions of this scenario are investigated. 
In both cases the baseline smooth initial conditions are chosen according to 
eqs.~(\ref{e:w}--\ref{e:all}). 
The hot-spots scenario refer to case where 
energy is superimposed on the smooth initial energy density profile. 
The hot-spots-with-momentum scenario refers to the case where energy as well as
momentum are superimposed onto the baseline initial conditions. 
Energy and momentum anisotropy is completely included in the initial conditions and not released over the finite time interval. 
In these regions, the same amount of energy and momentum is 
deposited as a hard parton would carry if it was produced
there. 
Also the number of hot spots corresponds to number of 
hard partons in simulations with hard partons. 
Their size is the same as would be the spatial spread of energy deposition from hard partons
at the beginning of the hydrodynamic evolution. 
Positions of hot spots are chosen from the distribution, which is
also the same as in the case with hard partons: \textit{i.e.}\ it is the distribution of initial binary collisions. 
Additional momentum deposition also modifies the initial expansion velocity field accordingly. 

Comparison in Fig. \ref{fig:fig1} shows the importance of momentum deposition 
\emph{during} the hydrodynamic evolution. 
It has been shown previously \cite{jets} that the hot-spots simulation does not produce 
the same amount of momentum anisotropy as the forces by which hard partons pull the 
plasma. Here, we  show that neither
fluctuations with momentum deposition in the initial conditions by themselves are able to generate the same flow anisotropies as 
wakes with streams induced by hard partons. 
Note that non-vanishing of $v_1$ is caused by transverse momentum of hard partons which have 
escaped the studied rapidity interval $\langle -1, 1\rangle$ and lead to the transverse momentum imbalance. 

\begin{figure}[h]
 \centering
  \includegraphics[width=0.65\textwidth]{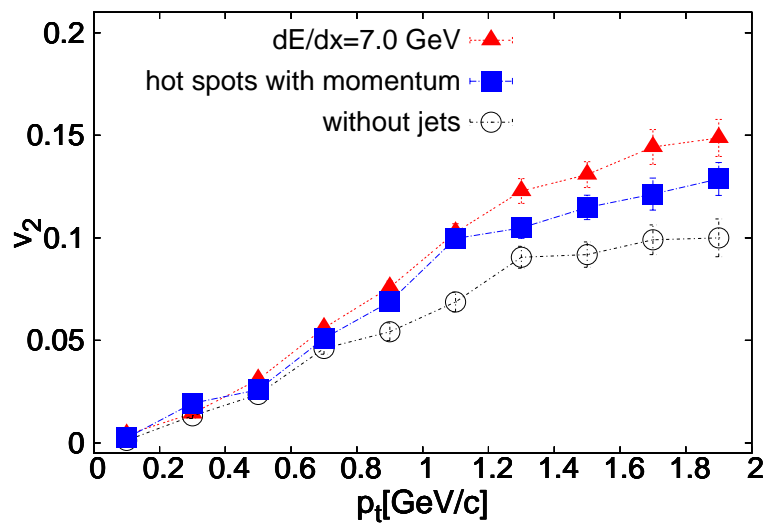}
  \caption{Anisotropy parameters $v_{2}$ for charged hadrons as functions of $p_{t}$
from collisions within centrality class 30-40\%. The energy loss of hard partons is given by 
$dE/dx|_0=7$~GeV/fm, other scenarios as in Fig.~\ref{fig:fig1}. }
\label{fig:v2} 
\end{figure}
\begin{figure}[h]
\centering
\includegraphics[width=0.65\textwidth]{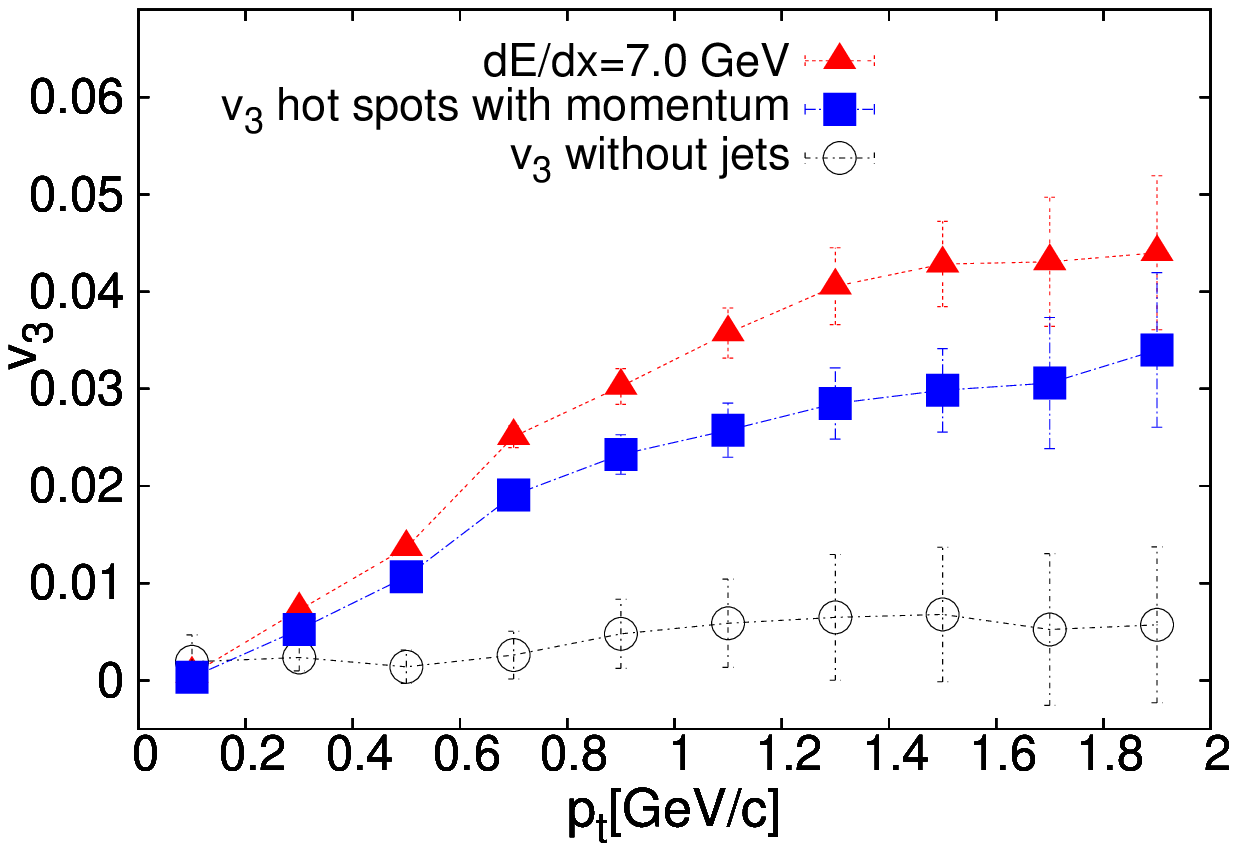}
\caption{Anisotropy parameters $v_{3}$ for charged hadrons as functions of $p_{t}$
from collisions within centrality class 30-40\%. The energy loss of hard partons is given by 
$dE/dx|_0=7$~GeV/fm, other curves as in Fig.~\ref{fig:v2}.}
\label{fig:v3} 
\end{figure}
Next, we move further in centrality and 
simulate sets of events within centrality class 30--40\% ($b=6$--7~fm). 
Results for anisotropy coefficient $v_{2}$ are shown in Fig. \ref{fig:v2}. 
Similarly, results for $v_{3}$ are presented in Fig. \ref{fig:v3}. 
The results of $v_{2}$ demonstrate that the flow 
anisotropy generated by hard partons is correlated with the reaction plane. 
With hard partons added, $v_2$ grows by about 50\% with respect to simulation with smooth
initial conditions and no hard partons. Of course, $v_3$ vanishes in the absence of 
hard partons or hot spots due to the lack of third order anisotropy within the bulk matter. 

Again here, we investigate whether the effect of energy-momentum deposition from 
hard partons can be represented by an appropriate choice of the initial conditions. In \cite{jets}
we have shown that adding hot spots with energy deposition does not lead to 
the same effect as hard partons. Here, 
in Figs.~\ref{fig:v2} and \ref{fig:v3} we show results of simulations 
with energy \emph{and momentum} superposition 
onto the smooth profile of the initial conditions. 
Also here, the amount and distribution of energy and momentum is the same as would 
be carried by hard partons if they were integrated into the evolution. We observe that even
the inclusion of hot spots with momentum into the initial state cannot  account for the whole 
effect on anisotropy which is generated by momentum deposition from hard partons during 
the evolution of the hydrodynamic bulk. 


\section{Summary and conclusions}
\label{s:summ}

Our results show that the momentum deposition during  fireball evolution 
must be included in a realistic hydrodynamic simulation.

As a consequence, 
the linear relation between initial state anisotropies and  $v_{n}$'s
may not be fully justified because in our suggested mechanism 
the anisotropy can rise during the fireball evolution. 
This is  most clearly shown in our results for ultra-central collisions (Fig.~\ref{fig:fig1}). 
Initial state anisotropies of all orders vanish in those simulations. 

We restricted our simulations here to the academic case with no fluctuations in the initial state 
of the hydrodynamic evolution. This allowed us to estimate the effect of momentum deposition 
from hard partons onto the resulting momentum anisotropies. The real quantitative influence can only 
be evaluated in  complete simulations with initial state fluctuations. It might well be that the observed data 
on $v_n$'s can be reproduced with suitably chosen initial conditions even if our mechanism is left out. 
However, the mechanism proposed here is quite natural and should be included. One should be careful in making 
conclusions about the initial state fluctuations based on simulations with our mechanism not included.  

A related question 
is then posed: can one find a relevant measure of spatial fireball anisotropy which is 
then translated into final-state momentum anisotropy? Perhaps the evolution of 
anisotropy decomposition into orthogonal terms 
\cite{Floerchinger:2013rya,Floerchinger:2013vua} may provide some 
insights here.

Even more promising tool might be including femtoscopy among the methods to analyse fireball dynamics. 
The $v_n$'s can be caused by an anisotropy of the fireball in spatial shape and/or expansion pattern
at the freeze-out. Both these effects cannot be determined uniquely just by analysis of single-particle 
distributions \cite{Tomasik:2004bn,Csanad:2008af,Lokos:2016fze}.
Fireball evolution with or without momentum deposition might end up in different spatial shapes even though
they produce the same $v_n$'s. 
Thus a complementary view might be obtained by looking at the azimuthal dependence of the femtoscopic 
correlation radii in second, third, and possibly higher orders \cite{Tomasik:2004bn,Csanad:2008af,Lokos:2016fze}.
Such an analysis has not been performed too often in experiments; only PHENIX collaboration has published results 
on third-order oscillations of correlation radii \cite{Adare:2014vax}. Perhaps one can even refrain from aligning 
the events to the second-order or the third-order event plane and use the method of Event Shape Sorting 
to select events so that oscillations to all orders can be studied together \cite{Kopecna:2015fwa}.

Figures \ref{fig:v2} and  \ref{fig:v3} confirm the results obtained from ultra-central sets 
of events, that the interplay 
of many hard-partons-induced streams causes significant contribution 
to the azimuthal anisotropies. This cannot be replaced by 
initial-state-generated flow anisotropies.
 
It remains to explore the influence of viscosities and fluctuating initial conditions on anisotropic parameters for all presented scenarios. 
The effect of shear viscosity is the drag in case of velocity gradients transverse to the direction of the velocity. Therefore, the momentum
which is transferred from the hard parton onto the fluid might spread into a broader stream. 
This is in line with the  simulations of Mach cones in the partonic transport model BAMPS which reasonably resembles 
viscous 3+1-dimensional hydrodynamics \cite{Bouras:2012mh,bouras}. 
The broader streams would be more likely 
to interact. Recall that it is the interaction and merging of such streams which is crucial for the effect of aligning the flow anisotropy 
with the geometry of the collision \cite{toymodel}. 
Hence, surprisingly, viscosity can even amplify the momentum anisotropy observed.

We have not investigated the effect of different path dependence of the energy loss. Changing the types
of path dependence (e.g.~linear, quadratic, \dots) would lead to changes of where the momentum is deposited 
during the hydrodynamic evolution. Since no visible variation of the resulting $v_n$'s was observed when varying the 
absolute size of $dE/dx$ within reasonable bounds, we do not expect much effect here either. The tool to distinguish 
various models might rather be the study of elliptic flow at high $p_t$ which comes from hard partons that survived 
and fly out of the medium.

The ultimate goal will be to provide unified description of the momentum anisotropies in both low-$p_t$ and high-$p_t$ 
regimes. Such an attempt has been made with the help of parton energy loss model JEWEL 
\cite{Zapp:2008af,Zapp:2013vla,Zapp:2013zya}. It has been combined with hydrodynamical model but only the influence 
on the radial flow has been estimated to be low \cite{Floerchinger:2014yqa,Zapp:2014msa}, 
while here we were interested in flow anisotropies. 
(We have checked that the azimuthal angle-integrated single-particle spectra are  hardly changed  in our simulations, also.) 
Thus a part of the task is also to find the overlap with other models and compare individual models on the market.

\paragraph{Acknowledgements}
This work was partially supported by RVO68407700, LG15001 (Czech Republic),
VEGA~1/0469/15 and APVV-0050-11 (Slovakia).

\end{document}